\documentclass[aps,pra,twocolumn,amsmath,amssymbs,superscriptaddress]{revtex4-1}

\usepackage{graphicx}
\usepackage{amsmath,amsfonts,amssymb}
\usepackage{epsfig}
\usepackage{wrapfig}
\usepackage{bbm}
\usepackage[usenames]{color}
\usepackage{array}
\usepackage{times}

\graphicspath{{Figures/}}

\def\wm{\omega_{m}}
\def\w{\omega}
\def\Vm{V_{m}}
\def\Vpi{V_{\pi}}

\def\w{\omega}
\def\t{t_{1}}
\def\t2{t_{2}}
\def\te{t_{e}}
\def\tl{t_{l}}

\def\delt{\delta t}

\def\lk{\left |}
\def\rk{\right >}

\def\ad{a^{\dag}}

\begin{document}

\title{Temporal and Spectral Manipulations of Correlated Photons using a Time-Lens}
\author{Sunil Mittal}
\author{Venkata Vikram Orre}
\affiliation{Joint Quantum Institute, NIST/University of Maryland, College Park, MD 20742, USA}
\affiliation{Department of Electrical and Computer Engineering, and IREAP, University of Maryland, College Park, MD 20742, USA}
\author{Alessandro Restelli}
\affiliation{Joint Quantum Institute, NIST/University of Maryland, College Park, MD 20742, USA}
\author{Reza Salem}
\affiliation{PicoLuz, LLC, Jessup, MD 20794, USA}
\author{Elizabeth A. Goldschmidt}
\affiliation{U.S. Army Research Laboratory, Adelphi, MD 20783, USA}
\affiliation{Joint Quantum Institute, NIST/University of Maryland, College Park, MD 20742, USA}
\author{Mohammad Hafezi}
\affiliation{Joint Quantum Institute, NIST/University of Maryland, College Park, MD 20742, USA}
\affiliation{Department of Electrical and Computer Engineering, and IREAP, University of Maryland, College Park, MD 20742, USA}


\begin{abstract}

A common challenge in quantum information processing with photons is the limited ability to manipulate and measure correlated states. An example is the inability to measure picosecond scale temporal correlations of a multi-photon state, given state-of-the-art detectors have a temporal resolution of about 100 ps. Here, we demonstrate temporal magnification of time-bin entangled two-photon states using a time-lens, and measure their temporal correlation function which is otherwise not accessible because of the limited temporal resolution of single photon detectors. Furthermore, we show that the time-lens maps temporal correlations of photons to frequency correlations and could be used to manipulate frequency-bin entangled photons. This demonstration opens a new avenue to manipulate and analyze spectral and temporal wavefunctions of many-photon states.

\end{abstract}

\maketitle

\section{Introduction}
Photons entangled in spectral-temporal degrees of freedom are extremely advantageous for robust, long-distance entanglement distribution \cite{Gisin2007, Pan2012, Brendel1999, Sasaki2011}. This characteristic feature has led to the development of a variety of techniques for spectral and temporal manipulations of single photons \cite{Vandevender2004, Tanzilli2005, Kolchin2008, Belthangady2009, Odele2015, Matsudae2016, Allgaier2017, Wright2017}. Recently, spectral compression of photons has gained widespread attention in order to efficiently interface wide-band sources of correlated photons with narrow-band nodes of a quantum network, for example, quantum dots and atomic systems \cite{Kimble2008, Lavoie2013, Karpinski2016, Allgaier2017}. At the same time, temporal magnification of photons facilitates high-fidelity photonic measurements in quantum simulations \cite{Walther2012, Carusotto2013, Hartmann2016, Noh2017}. For example, on-chip temporal boson-sampling and quantum walks \cite{Motes2014, He2014, Pant2016, Schreiber2012, Mittal2016} can have photonic wavepackets with temporal features shorter than the resolution of existing single photon detectors \cite{Marsili2013,Natarajan2012, Calandri2016}.

A versatile approach to spectrally compress and temporally magnify single photons is using time-lens techniques \cite{Kolner1994}. While time-lensing has been used widely in the past for temporal magnification of classical light pulses \cite{Bennett1999, Foster2008, Foster2009}, its use for single photons is very recent. Specifically, time-lens based techniques have demonstrated spectral manipulations of single photons \cite{Lavoie2013, Donohue2016, Karpinski2016} and also time-resolved detection of a single photon arriving in two time-bins \cite{Donohue2013}. However, these demonstrations have only manipulated single photons. It is highly desirable to manipulate and also measure temporal and spectral correlations of multi-photon states.
\begin{figure}[h!]
 \includegraphics[width=0.45\textwidth]{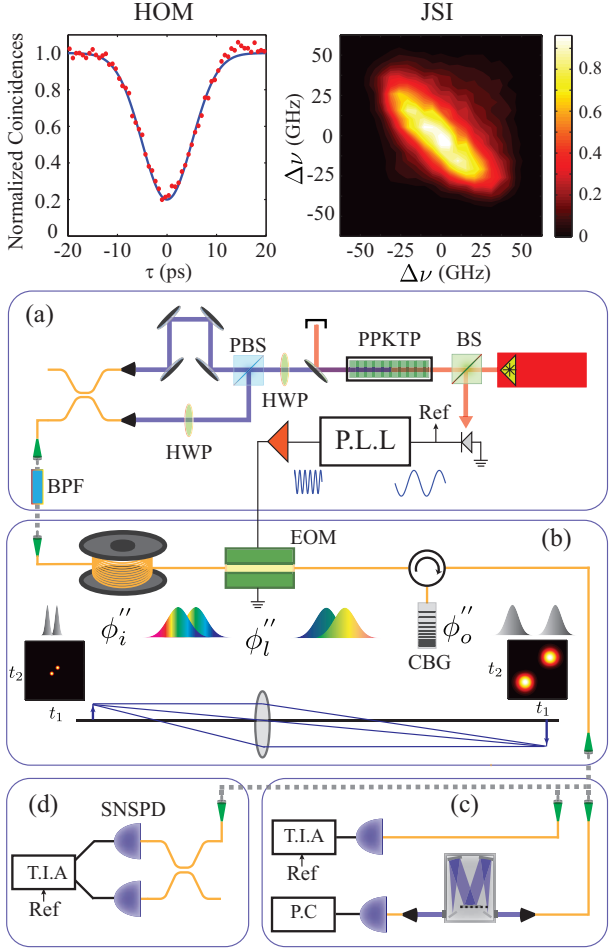}
  \vspace{1.0cm}
 \caption {
 \textbf{(a)} Time-bin entangled photons are generated using Type-II SPDC and a combination of a half-wave plate (HWP), a polarization beam-splitter (PBS) and a delay line. Insets show the measured HOM interference with a visibility of $\approx 80\%$ and nearly symmetric JSI, after the bandpass filter (BPF). The relative phase $\theta$ introduced by the delay line was stabilized using interference of another CW laser. \textbf{(b)} A converging time-lens is implemented using 15 km of SMF-28 fiber, an electro-optic phase modulator (EOM) and a chirped Bragg grating (CBG) which emulates 150 km of SMF-28 fiber. \textbf{(c)} A SNSPD and a TIA are used for single-channel time-resolved detection of photons and a monochromator along with a SNSPD are used for spectral measurements. \textbf{(d)} For JTI measurements, the output of the time-lens is fed to a fused-fiber beamsplitter connected to two SNSPDs and a time-tagged coincidence counting electronics.
 }
 \label{fig:1}
\end{figure}

In this work, we use an electro-optic phase modulator (EOM) based time-lens to magnify the two-photon temporal wavefunction associated with time-bin entangled photons while simultaneously preserving their quantum correlations. Our time-lens is designed to work in the telecom domain and achieves a temporal magnification of 9.6(2)x. First, we use this magnification to resolve two photons with a delay much less than the resolution of our superconducting nanowire single photon detectors (SNSPDs). Then, we measure joint-temporal intensity (JTI) of the magnified two-photon wavefunction, which is otherwise not measurable because of the limited detector resolution, and distinguish correlations between bunched and anti-bunched time-bin entangled photon pairs. Finally, we show that the time-lens maps temporal correlations of incoming photons to frequency correlations of outgoing photons and can be used to manipulate frequency-bin entangled two-photon states \cite{Ramelow2009}.

\section{Time-Lens Setup}

Fig.\ref{fig:1} illustrates a schematic of our time-lens setup. A dispersive element with a group delay dispersion (GDD) $\phi^{''}_{i} = \frac{d^{2}\phi_{i} \left(\omega \right)}{d\omega^2}$ is first used to spectrally chirp the input photon pulses. Here $\w$ is the angular frequency and $\phi_{i}\left(\w\right)$ is the frequency dependent phase-shift accumulated during propagation. A time-lens is then implemented using an electro-optic phase modulator (EOM) driven with a rf field of angular frequency $\omega_{m}$, amplitude $V_{m}$. It imposes a time-varying phase-shift $\phi_{l}(t) = -\frac{\pi V_{m}}{V_{\pi}}\cos\left(\omega_{m}t\right)$, where $V_{\pi}$ is the $\pi$ phase-shift voltage. When $ \wm t \ll 1$ and the time of arrival of photons is locked to the phase of the rf drive, the phase-shift can be approximated as $\phi_{l}(t) = \frac{\pi V_{m}}{2 V_{\pi}} \omega_{m}^{2} t^{2}$ with corresponding GDD $\phi^{''}_{l} = \frac{\Vpi}{\pi \Vm \wm^{2}}$. This quadratic time-varying phase introduced by the time-lens is exactly analogous to the spatially-varying phase imposed by a spatial lens. Furthermore, similar to a spatial lens which introduces transverse momentum shifts because of its curvature, the quadratic phase modulation and the associated GDD in a time-lens results in a linear frequency shift between two photons incident on the time-lens with a delay $\delta t_{\text{in}}$, given by
\begin{equation}\label{TL Frequency Shift}
  \delta \nu = \frac{\Vm}{\Vpi} \frac{\wm^{2}}{2} \delt_{\text{in}}.
\end{equation}
Therefore, the time-lens linearly maps the information contained in temporal degree of freedom of photons to the frequency domain. This is again analogous to the action of a spatial lens which Fourier transforms spatial information about an object to momentum domain. Finally, photons are subject to a large GDD at the output $\left(\phi^{''}_{o}\right)$ where the frequency shift $\delta \nu$ leads to a differential delay $ 2\pi \delta \nu \phi_{o}^{''} $.  The total delay between the photons at the output of the lens is
\begin{equation}\label{FrequencyToTimeMapping}
\delt_{\text{out}} = \delt_{\text{in}}  +   \frac{\pi V_{m}}{V_{\pi}} \wm^{2} \phi_{o}^{''} \delt_{\text{in}}.
\end{equation}
When the three dispersive elements satisfy the lens-equation \cite{Kolner1994}
\begin{equation}\label{LensEquation}
  -\frac{1}{\phi^{''}_{l}} = \frac{1}{\phi^{''}_{i}} + \frac{1}{\phi^{''}_{o}},
\end{equation}
the output is a temporally magnified image of the input with magnification $M =  \frac{\delt_{\text{out}}}{\delt_{\text{in}}} = -\frac{\phi^{''}_{o}}{\phi^{''}_{i}}$. The negative magnification implies that the time-lens creates temporally inverted image of the input photons. Note that similar to a spatial lens, a time-lens has a finite aperture $\tau_{a} \approx \frac{1}{\wm}$ and therefore, can only be used with pulsed light sources \cite{Kolner1994}.

Our experiment was designed to achieve a magnification of $\approx$9.8x. The initial GDD was introduced by 15 km spool of SMF-28 fiber with $\phi^{''}_{i} = -326 ~\text{ps}^{2}$. A large output GDD $\phi^{''}_{o} = -3190 ~\text{ps}^2$ corresponding to 150 km of SMF-28 was achieved by using a chirped Bragg grating (CBG). The EOM was driven by a rf signal with frequency $\nu_{m} = \frac{\omega_{m}}{2\pi} = 2.786$ GHz and was locked to the Ti-Sapphire laser. The $\pi$-phase-shift voltage $V_{\pi}$ of the modulator was measured to be 3.49(6) V, at 2.786 GHz. The rf signal amplitude $V_{m}$ was set to 12.3 V so that the GDD introduced by the EOM $\phi^{''}_{l}  \approx296 ~\text{ps}^2$ and satisfies the time-lens equation. Note that the GDD introduced by the lens is normal (positive) whereas that of input and output fibers is anomalous (negative). With these conditions, the lens is a converging lens \cite{Kolner1994}.

\section{Results}

\subsection{Temporal Magnification}

\begin{figure}
 \centering
 \includegraphics[width=0.48\textwidth]{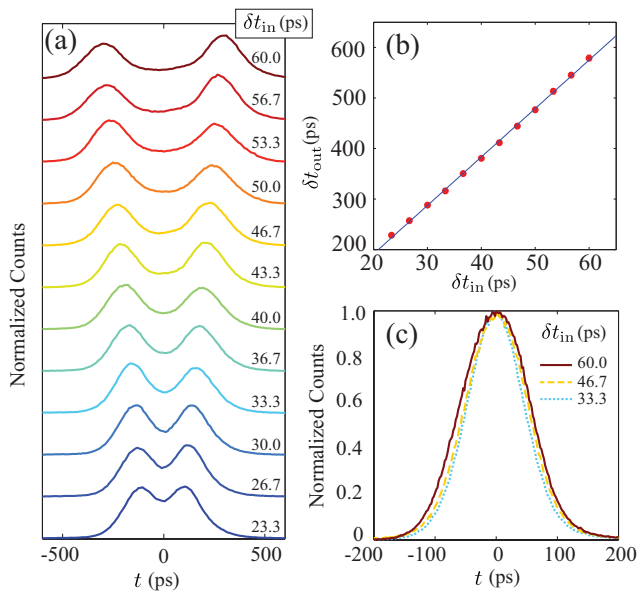}
 \caption {
 \textbf{(a)} Observed photon pulses after the time-lens, for different input time delays $\delt_{\text{in}}$. The two photons are very well resolved after the time lens for delay as small as $\approx$23 ps. \textbf{(b)} Measured delay (red markers) between photons at the output of the lens as a function of delay at the input. The delay increases linearly, with a slope $M = 9.6(2)$ where the uncertainty is from the linear fit (blue solid line). The size of the errorbars, representing statistical error in finding peaks of photon pulses, is less than the size of markers.  \textbf{(c)} Because of the detector jitter ($\approx ~100$ ps) of SNSPD, without the time-lens, the two photons cannot be resolved even for delay $\delt_{\text{in}}$ as large as 60 ps.
 }
 \label{fig:2}
\end{figure}

To demonstrate the working and resolving power of our time-lens, we first injected two photons into the lens, one arriving in early time-bin $\te$ and the other arriving in late time-bin $\tl$. The delay between the two time bins $\delt_{\text{in}} = \tl - \te$ was tunable and was chosen to be 20 - 60 ps, smaller than the timing jitter $\left(106~\text{ps} \right)$ of the detector so that the two photons cannot be directly resolved. The two photons were generated using a Type-II, collinear spontaneous parametric down conversion (SPDC) process in a 30 mm periodically-poled KTP crystal pumped by a pulsed Ti-Sapphire laser emitting $\approx$10 ps pulses at $\approx$ 775.45 nm wavelength (see Fig.\ref{fig:1}(a)). The crystal was phase-matched to produce nearly degenerate, orthogonally polarized $(H ~\text{and} ~ V)$ signal and idler photons near 1550.9 nm, at $30^{0}$C. These orthogonally polarized photons were separated using a polarization beam splitter (PBS) and a relative delay was introduced between them. The $V$ polarized photons were converted to $H$ polarized using a half-wave plate (HWP) and then the photons were recombined into a single-mode fiber using a fused-fiber beamsplitter. The photons were subsequently filtered with a FWHM of $\approx75$ GHz (0.6 nm) and sent to the time lens. The lower bound on the photon pulsewidth was estimated to be 16.7(7) ps using HOM interference. The photons at the output of the time lens were detected using a superconducting nanowire single photon detector (SNSPD) and their arrival time was recorded using a Time Interval Analyzer (TIA) (see Fig.\ref{fig:1}(c)).

\begin{figure*}
 \centering
 \includegraphics[width=0.96\textwidth]{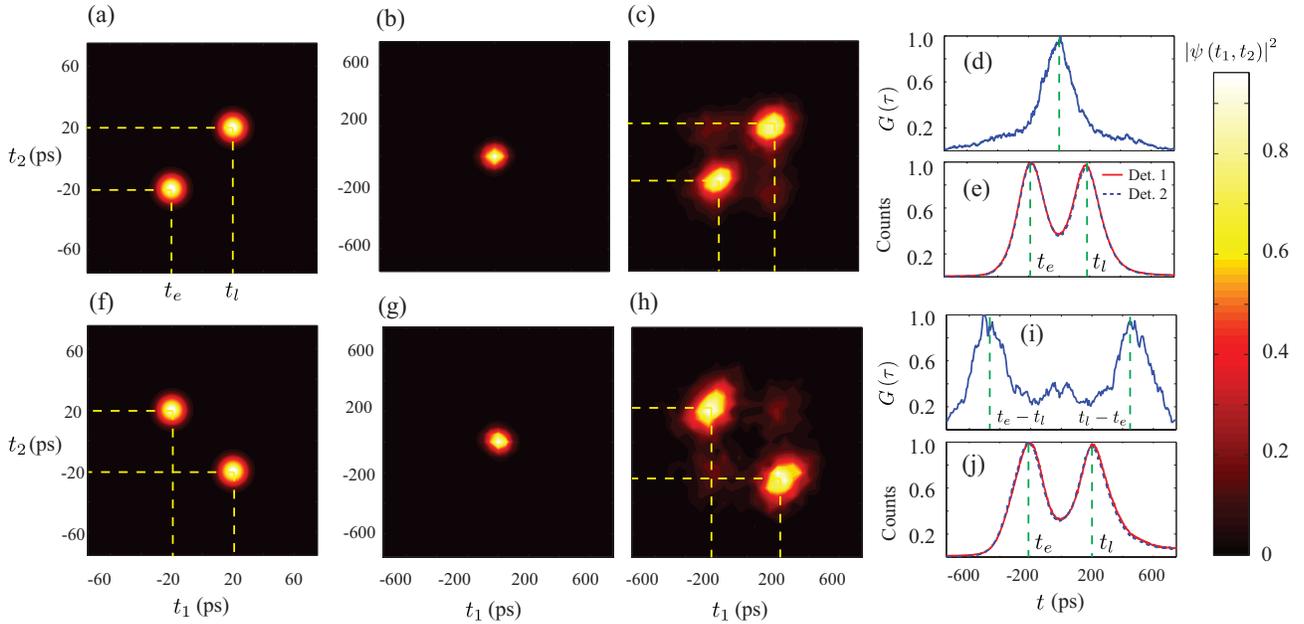}
 \caption {
 \textbf{(a)} Simulated JTI of the bunched two-photons state before the time-lens. \textbf{(b)} Measured JTI, without the time-lens. The temporal correlations cannot be resolved at all. \textbf{(c)} Measured JTI with the time-lens. The two photons can now be very clearly resolved, showing bunched behavior. \textbf{(d)} Measured $G\left(\tau\right)$ peaks at $\tau = 0$, consistent with bunched behavior. \textbf{(e)} Measured singles counts on two detectors. \textbf{(f)-(j)} Corresponding results for the anti-bunched state. $G\left(\tau\right)$ now peaks at $\tau = \pm \left(\te - \tl\right) \approx 420$ ps, showing anti-bunched photons.  Note that the single channel measurements of photon pulses cannot distinguish between the two states.
 }
 \label{fig:3}
\end{figure*}

Fig.\ref{fig:2}(a) shows the observed photon pulses at the output of the lens for different input delays $\delt_{\text{in}}$ between the two photons. We can clearly resolve the two photons with input delay as short as 23 ps, consistent with the estimated time resolution, ratio of the effective focal length to the aperture of the lens, $\delta t_{0} = \frac{2 V_{\pi}}{V_{m} \omega_{m}} \approx 30~\text{ps}$ \cite{Kolner1994}. Fig.\ref{fig:2}(b) plots the measured delay between photons at the output of the lens as a function of delay at the input. The slope of this linear plot is the magnification factor $M$, measured to be $9.6(2)$, in a very good agreement with the design value of $M = 9.8$. The high-fidelity of the time-lens is evident from the linearity of the plot which shows that the magnification is same throughout the lens aperture. The small discrepancy between the observed and the designed magnification factors is due to marginal overfilling of the time-lens aperture for higher $\delt_{\text{in}}$. Furthermore, the measured individual photon pulsewidth (FWHM) at the lens output is 186(1) ps (after correcting for detector jitter), in a good agreement with the observed magnification factor, given the input pulse width was estimated to be 16.7(7) ps. For comparison, Fig.\ref{fig:2}(c) shows the observed TIA response when the photons are incident on the detector without a time-lens and the two photons are completely unresolved by the detector.

\subsection{Measurement of Temporal Correlations}

Simple measurements of the time delay between two photons, which are essentially projective measurements of the two-photon temporal wavefunction, do not provide any insight into quantum correlations. For example, single channel delay measurements cannot distinguish between two-photon states corresponding to temporally bunched and anti-bunched photons \cite{Tittel1998}. In the bunched state $\left(\left|2_{e}, 0_{l}\right> - \left|0_{e}, 2_{l} \right> \right)$, both the photons arrive in the {\it early} time-bin or both in the {\it late} time-bin. In the anti-bunched state $\left( \left|1_{e}, 1_{l}\right> \right)$, one photon arrives early and the other late. An alternative is to measure the Joint-temporal intensity (JTI) which can characterize temporal correlations of a two-photon state, analogous to the joint-spectral intensity (JSI) which is used to characterize spectral correlations between photon pairs \cite{Harder2013}. JTI is the probability of finding two photons, one at time $t_{1}$ and the other at $t_{2}$, and is defined as $\left|\psi\left(t_{1},t_{2}\right) \right|^{2}$  where $\psi\left(t_{1},t_{2}\right)$ is the two-photon temporal wavefunction. Even though a JTI measurement does not measure the phase associated with the two-photon wavefunction, it is well suited for many quantum simulation techniques, for example, quantum walks and boson sampling, which require a measurement only of intensity correlations. JTI of a two-photon state can be easily measured using a beam-splitter and time-resolved coincident detection events at two detectors (see Fig.\ref{fig:1}(d)). However, direct JTI measurements are limited in time-resolution because of the detector jitter. While time-resolved frequency upconversion \cite{Kuzucu2008} and intensity modulation \cite{Belthangady2009} schemes allow JTI measurements with picosecond resolution by effectively introducing narrow filters in time or frequency, they require a two-dimensional scan of the filter position(s) for a two-photon state and therefore, can be extremely slow. In the following, we demonstrate that a time-lens expands the two-photon temporal wavefunction while preserving the quantum correlations of the wavefunction. This magnification allows us to directly measure the JTI, without any filtering, and hence unravel correlations of two-photon states with a resolution beyond the limitations imposed by detector jitter.

To generate two-photon states with bunched and anti-bunched temporal correlations, we use another HWP after the SPDC. When the HWP is set at an angle of $22.5^{0}$ with respect to the horizontal, it acts as a 50:50 beam-splitter for the $H$ and $V$ polarized photons. Furthermore, as shown in Refs.\cite{Kuo2016, Gerrits2015} and Appendix B, when the two-photon spectral wavefunction after the SPDC is symmetric with respect to exchange of photons, the two-photon state after the HWP is polarization entangled, i.e., $\left|2_{H}, 0_{V}\right> - \left|0_{H}, 2_{V} \right>$. The PBS and the delay line following the HWP map this polarization entangled state to the time-bin entangled state
\begin{eqnarray}\label{PsiB}
\nonumber \lk \Psi_{\text{B}} \rk = \int \int dt_{1} dt_{2} && \psi\left(t_{1},t_{2}\right)  [\ad\left(t_{1}-\te\right) \ad\left(t_{2}-\te\right) \\
                         && + e^{i\theta} \ad\left(t_{1}-\tl\right) \ad\left(t_{2}-\tl\right)]  \lk 0 \rk,
\end{eqnarray}
where $\ad\left(t-t_{e(l)}\right)$ is the photon creation operator corresponding to the {\it early (late)} time-bin and $\theta$ is the phase resulting from delay $\delta t_{\text{in}}$. This is a time-bin entangled two-photon state where the two photons are always bunched (B), either appearing in the {\it early} time-bin $\left(\te \right)$ or in the {\it late} time-bin $\left(\tl \right)$. Fig.\ref{fig:3}(a) shows the simulated JTI for this state, with the individual photon pulses assumed to be gaussian. In our experiment, the exchange symmetry of the two-photon spectral wavefunction was confirmed using high visibility ($\approx 80\%$) HOM interference and a direct measurement of the JSI of the two photons using chirped Bragg grating as a frequency-to-time converter (Fig.\ref{fig:1}) \cite{Harder2013, Gerrits2015}.

When the HWP angle is set to $0^{0}$, it does not mix the $H$ and $V$ polarized photons and therefore, the two-photon state at the input of the lens is
\begin{equation}\label{PsiAB}
 \lk \Psi_{\text{AB}} \rk = \int \int dt_{1} dt_{2} \psi\left(t_{1},t_{2}\right) \ad\left(t_{1}-\te\right) \ad\left(t_{2}-\tl\right)  \lk 0 \rk.
\end{equation}
Now, the two photons are anti-bunched (AB), i.e., they always arrive in different time-bins. Note that this state is not time-bin entangled but the beamsplitter used for JTI measurement after the lens cannot distinguish between the two photons and therefore, induces entanglement (see Appendix B). The simulated JTI for this anti-bunched state is shown in Fig.\ref{fig:3}(f).

Fig.\ref{fig:3}(b,c) show the measured JTI for the bunched state $\lk \Psi_{\text{B}} \rk$, without and with a time-lens, respectively. In the absence of time-lens, direct measurement of JTI (using setup shown in Fig.\ref{fig:1}(d)) cannot resolve any correlations in the two-photons state because the time-bins are separated by a delay $\left(40~\text{ps}\right)$ less than the timing jitter $\left(\approx100~\text{ps}\right)$ of the two detectors. By using a time-lens, we magnify the temporal correlations between the photons which are now easily resolved by JTI measurements (Fig.\ref{fig:3}(c)). A good agreement of the measured JTI with the simulated JTI shows that the time-lens faithfully magnifies the two-photon wavefunction while preserving its temporal correlations. A small probability of photons arriving in different time-bins (anti-bunched, along the anti-diagonal) is also observed in this plot. This is mainly because of multi-photon processes in the SPDC. The measured delay between the time-bins $\delt \approx 360~\text{ps}$ is consistent with the observed magnification.

To further quantify this behavior, in Fig.\ref{fig:3}(d) we plot the probability $G\left(\tau\right)$ of photons arriving with a time difference $\tau$, i.e.,
\begin{equation}
G\left(\tau\right) = \int\int dt_{1}dt_{2} \left|\Psi\left(t_{1},t_{2}\right)\right|^{2} \delta\left(\tau - t_{1} + t_{2}\right).
\end{equation}
As can be seen, $G\left(\tau\right)$ peaks at $\tau = 0$ again verifying that the photons are bunched.

Fig.\ref{fig:3}(f-j) show the corresponding results for the anti-bunched state $\lk \Psi_{\text{AB}} \rk$. Again, without the time-lens no correlations are observed in the JTI whereas with the time-lens we clearly see that the two photons always arrive in different time-bins.  The probability $G\left(\tau\right)$ now peaks at $\tau \approx \tl - \te$. Also, a finite probability of bunching (along the diagonal) is observed which is due to multi-photon processes in the SPDC. To further highlight the significance of JTI measurements, in Fig.\ref{fig:3}(e) and (j), we plot the observed singles count on the two detectors, for bunched and anti-bunched cases, respectively. The plots for the two states are exactly identical and have no information about their correlations. This confirms that single channel delay measurements, in general, cannot be used to characterize two-photon states.

\subsection{Measurement of Spectral Correlations}

\begin{figure}
\centering
 \includegraphics[width=0.48\textwidth]{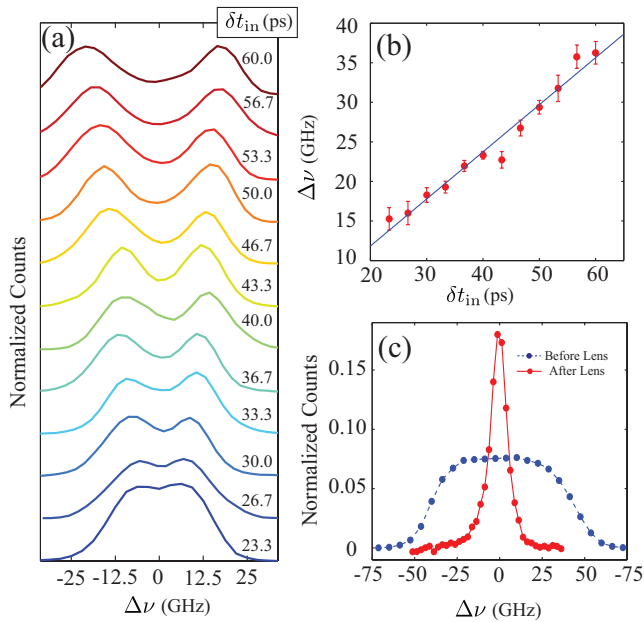}
 \caption {
  \textbf{(a)} Measured spectrum of photon pulses, for different input delay $\delt_{\text{in}}$. \textbf{(b)} Relative frequency shift as a function of input delay $\delt_{\text{in}}$. A linear fit (blue line) to the measured data (red markers) gives a slope of $0.60(8)$ and agrees well with slope of $0.54$ estimated using (\ref{TL Frequency Shift}). \textbf{(c)} Measured spectral profile before the time-lens (FWHM $\sim$ 75 GHz) and after the time-lens (FWHM $\sim$ 9 Ghz, corrected for monochromator bandwidth of 8.2GHz) gives a spectral compression factor of $\approx 8.3$x.
 }
 \label{fig:4}
\end{figure}

Now, we show that a time-lens also maps temporal correlations of input photons to frequency correlations of outgoing photons. As shown in (\ref{TL Frequency Shift}), the EOM introduces a frequency shift $\delta \nu$ between two photons separated by a temporal delay $\delt_{\text{in}}$ at its input. The CBG used after the EOM maps this frequency shift to time which is then measured using the TIA. Because this frequency-to-time mapping is linear, the time-axis in Fig.\ref{fig:3}(c,h) could be easily rescaled to frequency using (\ref{FrequencyToTimeMapping}) and shows that the two-photon wavefunction at the lens output is also frequency-bin entangled. To independently verify this frequency shift, we used a monochromator to measure the spectrum of photons at the lens output. Fig.\ref{fig:4}(a) shows the measured spectrum for different input delays $\delt$ and Fig.\ref{fig:4}(b) plots the frequency shift as a function of delay $\delt$. As expected, frequency shift increases linearly with a slope $0.60(8)$ which compares well with the slope $0.54$ estimated using (\ref{TL Frequency Shift}). We also confirmed spectral compression of single photons and Fig.\ref{fig:4}(c) plots the measured single-photon spectrum before and after the time-lens. The measured bandwidth is $\approx 75~\text{GHz}$ before the lens and $9(1)~\text{GHz}$ after the lens, corresponding to a spectral compression of $\approx 8.3$x.

\subsection{Coherence of the Time-Lens}

\linespread{1.0}
\begin{figure*}
 \centering
 \includegraphics[width=0.9\textwidth]{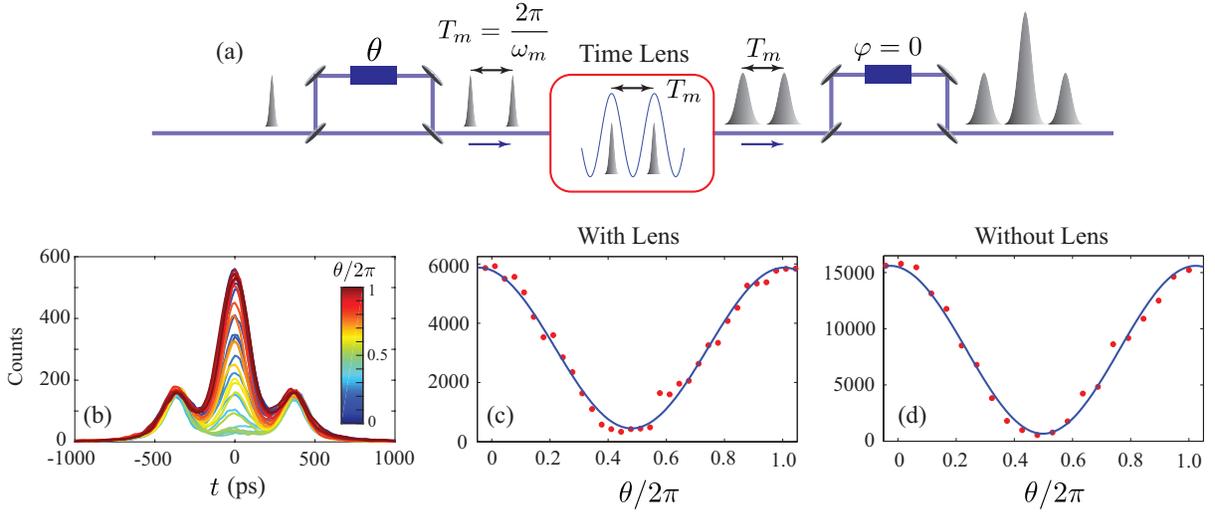}
 \caption{(a), Setup used to test coherence of the time-lens. A single photon is prepared in the superposition of \textit{early} and \textit{late} time-bins, with a phase $\theta$. The time difference between the early and late time bins is set to be equal to the time-period of the RF drive ($\approx$ 360 ps) for the EOM. In this configuration, the two time bins essentially see two time-lenses and do not acquire a relative frequency shift. At the output of the time-lens, a Franson interferometer, with same delay as the input interferometer, is used to measure the phase $\theta$. (b), Observed signal at the output for different values of phase $\theta$. (c)  Intensity of the middle peak as a function of phase $\theta$. The intensity varies as $\cos\left(\theta - \varphi\right)$, where $\varphi$ is the phase associated with the output interferometer. (d), Intensity of the middle peak as a function of $\theta$, without a time-lens.}
 \label{fig:5}
\end{figure*}
\linespread{2.0}

As shown in eq.(\ref{PsiB}), the \textit{early} and \textit{late} time-bins are associated with a relative phase $\theta$ arising from the delay in the input interferometer. A high-fidelity time-lens is expected to be coherent and preserve this relative phase. However, JTI measurements are insensitive to this relative phase and therefore, do not show the coherence of the time-lens. Furthermore, as shown in Fig.\ref{fig:4}, photons in the \textit{early} and \textit{late} time-bins acquire a relative frequency shift as they propagate through the time-lens. Therefore, the standard state tomography procedure using Franson interferometers can not measure this relative phase $\theta$ after the time-lens. For the same reason, the fidelity of the two-photon state after the time-lens can not be accessed using HOM interference.

Nevertheless, to show the coherence of the time-lens, we  prepare a single photon in a superposition state of \textit{early} and \textit{late} time-bins with relative phase $\theta$, $\left|e\right> + e^{i \theta} \left|l \right>$, such that the delay between the two time-bins is equal to the time-period $\left(\approx 360 \text{ps} \right)$ of the rf drive for the EOM (Fig.\ref{fig:5} (a)). With this arrangement, the two time-bins get magnified using two separate time-lenses but there is no relative frequency shift between the time-bins. Therefore, a Franson interferometer, with same delay as the input, can be used to measure the relative phase $\theta$ after the time-lens.

Fig.\ref{fig:5}(b) shows the measured temporal response at the output of the interferometer, for different values of phase $\theta$ and fixed phase $\varphi = 0$ of the output interferometer. The middle peak corresponds to interference of \textit{early} photons taking a longer path and \textit{late} photons taking a shorter path in the output interferometer. Fig.\ref{fig:5}(c) shows the intensity of this middle peak as a function of input phase $\theta$. As expected, its intensity is proportional to $\cos\left(\theta - \varphi \right)$. The visibility of this interference fringe was measured to be $\approx 86\%$. Furthermore, using a large delay of 360 ps between the two time-bins allows time-resolved detection of photons and therefore, a measurement of phase $\theta$ without the time-lens. Fig.\ref{fig:5}(d) shows the measured interference fringe with out the time-lens, with a visibility of $\approx 93\%$. The marginal reduction in interference visibility while using a time-lens is mainly because of the temporal magnification of photons which reduces the orthogonality between the two time-bins. This observation of high visibility single photon interference at the output of the time-lens clearly demonstrates that the temporal magnification is coherent and preserves the relative phase between \textit{early} and \textit{late} time-bins.

\section{Conclusion}
We have shown $9.6$x temporal magnification of a two-photon temporal wavefunction using a deterministic, electro-optic modulator based time-lens. In this demonstration, the time-lens was driven at only $2.8$ GHz whereas commercially available EOMs can easily achieve 40 GHz operation. By using higher rf frequencies, this technique could easily be adapted to achieve much higher magnification and picosecond scale temporal resolution, using existing single photon detectors. Furthermore, we used a two-photon source entangled in two discrete time bins. However, our scheme is more general and can be used to measure arbitrary temporal correlations of multi-photon states, for example, those arising from temporal quantum walks.

\appendix

\section{Action of a Time-Lens on a Two-Photon Wavefunction}

In this section, we derive the relations governing the action of a time-lens on a two-photon wavefunction. We start with a general two-photon state at the input of the lens
\begin{equation}
 \lk \Psi \rk = \int \int dt_{1} dt_{2} \psi_{\text{in}}\left(t_{1},t_{2}\right) \ad\left(t_{1}\right) \ad\left(t_{2}\right) \lk 0 \rk,
\end{equation}
where $\psi_{\text{in}}\left(t_{1},t_{2}\right)$ is the two-photon temporal wavefunction. Using 2D Fourier transform, the temporal wavefunction could be written in frequency domain as
\begin{equation}
  \psi_{\text{in}}\left(t_{1},t_{2}\right) = \frac{1}{2\pi} \int \int d\w_{1} d\w_{2} e^{i\w_{1} t_{1}}  e^{i\w_{2} t_{2}} \tilde{\psi}_{\text{in}} \left(\w_{1},\w_{2}\right),
\end{equation}
where $\tilde{\psi}_{\text{in}} \left(\w_{1},\w_{2}\right)$ is now the two-photon spectral wavefunction at the input.

Following Ref. \cite{Kolner1994}, this two-photon state is first subject to an input group delay dispersion $\phi_{i}^{''}$ which results in a chirped temporal wavefunction $\left(\psi_{\text{ch}}\left(t_{1},t_{2}\right) \right)$ given as
\begin{eqnarray}
\nonumber  \psi_{\text{ch}} &&\left(t_{1},t_{2}\right) = \frac{1}{2\pi} \int \int d\w_{1} d\w_{2} e^{i\w_{1} t_{1}}  e^{i\w_{2} t_{2}} \\
 && e^{\left(-i \phi_{i}^{''}\frac{\left(\w_{1}-\w_{0}\right)^{2}}{2}\right)} e^{\left(-i \phi_{i}^{''}\frac{\left(\w_{2}-\w_{0}\right)^{2}}{2} \right)} \tilde{\psi}_{\text{in}} \left(\w_{1},\w_{2}\right).
\end{eqnarray}
Here $\w_{0}$ is the central frequency of the spectral wavefunction.

After the input dispersion, the chirped two-photon wavefunction enters the EOM. The EOM adds a time-dependent phase $\phi_{l}\left(t \right) = -\frac{\pi \Vm}{\Vpi} \cos\left(\wm t \right)$ to the wavefunction such that the two-photon wavefunction after the EOM is given as
\begin{equation}\label{EOM_Temporal}
  \psi_{\text{EOM}}\left(t_{1},t_{2}\right) =   e^{\left(i \frac{\pi \Vm}{2 \Vpi} \wm^{2} t^{2}_{1} \right)}  e^{\left(i \frac{\pi \Vm}{2 \Vpi} \wm^2 t^{2}_{2} \right)} \psi_{\text{ch}} \left(t_{1},t_{2}\right).
\end{equation}

Finally, photons are subject to a large GDD at the output $\left(\phi^{''}_{o}\right)$ which acts as a frequency-to-time converter and the temporal wavefunction at the output of the time-lens is
\begin{eqnarray}
\nonumber  \psi_{\text{out}}&& \left(t_{1},t_{2}\right) = \frac{1}{2\pi} \int \int d\w_{1} d\w_{2} \exp(i\w_{1} t_{1}) \exp(i\w_{2} t_{2})  \\
&& e^{\left(-i \phi_{o}^{''}\frac{\left(\w_{1}-\w_{0}\right)^{2}}{2} \right)} e^{\left(-i \phi_{o}^{''}\frac{\left(\w_{2}-\w_{0}\right)^{2}}{2} \right)} \tilde{\psi}_{\text{EOM}}\left(\w_{1},\w_{2}\right),
\end{eqnarray}
where $\tilde{\psi}_{\text{EOM}}\left(\w_{1},\w_{2}\right)$ is the Fourier transform of $\psi_{\text{EOM}}\left(t_{1},t_{2}\right)$. Using above equations, the temporal wavefunction at the output of the lens can be easily calculated for any general two-photon wavefunction at its input.

\section{Generation of Time-Bin Entangled Two-Photon States}

In this section we discuss the formalism to generate of time-bin entangled photons using a combination of a HWP, a PBS and a delay line. We start with writing the two-photon state just after the SPDC as
\begin{eqnarray}
\nonumber  \lk \Psi \rk &=& \int\int d\w_{1} d\w_{2} \tilde{\psi}\left(\w_{1},\w_{2}\right) \ad_{H}\left(\w_{1}\right) \ad_{V}\left(\w_{2}\right) \lk 0 \rk \\
                  &=& \int \int dt_{1} dt_{2} \psi\left(t_{1},t_{2}\right) \ad_{H}\left(t_{1}\right) \ad_{V}\left(t_{2}\right) \lk 0 \rk,
\end{eqnarray}
where the temporal, $\psi\left(t_{1},t_{2}\right)$, and spectral, $\tilde{\psi}\left(\w_{1},\w_{2}\right)$, two-photon wavefunctions are related by the 2D Fourier transform.

Following SPDC, the two photons are subjected to a variable beamsplitter implemented using a HWP and a PBS. We first analyze the case when the HWP is oriented at an angle of $22.5^{0}$ with respect to the horizontal axis and results in a time-bin entangled state where the photons are always bunched (eq.\ref{PsiB}). The HWP acts as a $50:50$ beam splitter for the $H$ and $V$ polarized photons and leads to the two-photon state
\begin{eqnarray}
\nonumber  && \lk \Psi \rk = \int \int d\w_{1} d\w_{2} \tilde{\psi}\left(\w_{1},\w_{2}\right) [ \ad_{H}\left(\w_{1}\right) \ad_{H}\left(\w_{2}\right) - \\
&& \ad_{V}\left(\w_{1}\right) \ad_{V}\left(\w_{2}\right) + \ad_{V}\left(\w_{1}\right) \ad_{H}\left(\w_{2}\right) - \ad_{H}\left(\w_{1}\right) \ad_{V}\left(\w_{2}\right) ] \lk 0 \rk.
\end{eqnarray}

When the two-photon spectral wavefunction associated with the SPDC process is symmetric, i.e., $\tilde{\psi}\left(\w_{1},\w_{2}\right) = \tilde{\psi}\left(\w_{2},\w_{1}\right)$, the last two terms in the above expression cancel each other and the two-photon state is simply \cite{Gerrits2015,Kuo2016}
\begin{eqnarray}
 \nonumber \lk \Psi \rk = \int \int d\w_{1} d\w_{2} \tilde{\psi}\left(\w_{1},\w_{2}\right) && [ \ad_{H}\left(\w_{1}\right) \ad_{H}\left(\w_{2}\right) - \\
 && \ad_{V}\left(\w_{1}\right) \ad_{V}\left(\w_{2}\right)] \lk 0 \rk.
\end{eqnarray}
This is a polarization entangled state of two-photons where both the photons are either $H$ polarized or  $V$ polarized. This phenomenon is similar to the usual HOM interference with a beamsplitter where both the photons at the output of the beamsplitter go into same port \cite{Gerrits2015,Kuo2016}. Here, the two polarization modes $H$ and $V$ are analogous to the two spatial modes, and the HWP works as the beam splitter.

To map this polarization entanglement to time-bin entanglement, we use a polarization beam splitter (PBS) to separate the $H$ and $V$ polarized photons. We then introduce a relative delay, $\delt_{\text{in}} = t_{l} - t_{e}$, between the two paths such that $H$ polarization corresponds to the early time-bin $\te$ and $V$ to the late time-bin $\tl$. Another HWP is then used to convert the $V$ polarized photons to $H$. Subsequently, photons from both the arms are collected in two PMFs and combined using a fused fiber beam-splitter. The two-photon state after the fiber beamsplitter is
\begin{eqnarray}\label{Phase}
\nonumber  \lk \Psi \rk &=& \int \int d\w_{1} d\w_{2} \tilde{\psi}\left(\w_{1},\w_{2}\right) [ e^{-i\w_{1} \te} e^{-i\w_{2} \te} \ad\left(\w_{1}\right) \ad\left(\w_{2}\right) \\
\nonumber && - e^{-i\w_{1} \tl} e^{-i\w_{2} \tl} \ad\left(\w_{1}\right) \ad\left(\w_{2}\right)] \lk 0 \rk \\
\nonumber  &=& \int \int dt_{1} dt_{2} \psi\left(t_{1},t_{2}; \te,\tl\right) [ \ad\left(t_{1}-\te\right) \ad\left(t_{2}-\te\right) \\
  && - \ad\left(t_{1}-\tl\right) \ad\left(t_{2}-\tl\right)] \lk 0 \rk.
\end{eqnarray}
This is a time-bin entangled two-photon state where the photons always arrive bunched, either at time $\te$ or at time $\tl$. We have dropped polarization indices in this state because now both the photons always $H$ polarized. The JTI for this state is shown in Fig.\ref{fig:3}(a).

Next, we analyze the case when the HWP is set at an angle of $0^{0}$, i.e., its axis is aligned with the horizontal and leads to generation of anti-bunched two-photon state (eq.\ref{PsiAB}). With this setting, the HWP does not rotate the polarizations of the two photons and therefore, the two-photon state after the HWP is the same as that generated by the SPDC. It imprints an overall $\pi$ phase on the two-photon wavefunction which is inconsequential. As before, we associate $H$ and $V$ polarized photons with \textit{early} and \textit{late} time bins and, the two-photon state at the output of the beam-splitter is
\begin{equation}
\nonumber  \lk \Psi \rk = \int \int dt_{1} dt_{2} \psi\left(t_{1}-\te,t_{2}-\tl\right) \ad\left(t_{1}-\te\right) \ad\left(t_{2}-\tl\right) \lk 0 \rk.
\end{equation}
Note that this state is not a time-bin entangled state. It is simply a correlated, separable state of two photons where one comes early and the other late. However, for JTI measurements, we use another fiber beamsplitter after the time-lens. The two output ports of the beamsplitter are connected to single photon detectors each. The two-photon state after the beamsplitters is given as
\begin{eqnarray}
\nonumber && \lk \Psi \rk = \int \int dt_{1} dt_{2} \psi\left(t_{1}-\te,t_{2}-\tl\right)  \\
&& \left(d_{1}^{\dagger}\left(t_{1}-\te\right) -i d_{2}^{\dagger}\left(t_{1}-\te\right) \right) \left(d_{1}^{\dagger}\left(t_{2}-\tl\right) -i d_{2}^{\dagger}\left(t_{2}-\tl\right) \right) \lk 0 \rk.
\end{eqnarray}
where $d_{1,2}^{\dagger}$ are the photon creation operators on detectors $1$ and $2$. A measurement of the coincident events on two detectors then projects this state to
\begin{eqnarray}
\nonumber && \lk \Psi \rk = \int \int dt_{1} dt_{2} \psi\left(t_{1}-\te,t_{2}-\tl\right) \\
 && \left(d_{1}^{\dagger}\left(t_{1}-\te\right) d_{2}^{\dagger}\left(t_{2}-\tl\right)+ d_{2}^{\dagger}\left(t_{1}-\te\right) d_{1}^{\dagger}\left(t_{2}-\tl\right) \right) \lk 0 \rk.
\end{eqnarray}
This is a measurement induced entangled state where the two photons are always anti-bunched. When detector-1 records and {\it early} event at time $\te$, detector-2 would record a {\it late} event at time $\tl$ and vice-versa. The simulated JTI for this state is shown in Fig.\ref{fig:3}(f).

\section{Contribution of Multi-Photon Processes to Measured JTI}

In the experimentally measured JTI (Fig.\ref{fig:3}(c)) we observe some anti-bunched photons at the lens output for ideally bunched photons at the input and vice-versa. These photon pairs with opposite correlations than expected are because of multi-photon processes in the SPDC. To estimate this contribution, we begin with approximating the multi-photon state after the SPDC as \cite{MilburnBook}
\begin{equation}
  \lk \Psi \rk = \sqrt{(1- p_{1} - p_{2})} \lk 0_{H}, 0_{V} \rk +  \sqrt{p_{1}} \lk 1_{H}, 1_{V} \rk + \sqrt{p_{2}} \lk 2_{H}, 2_{V} \rk,
\end{equation}
where $p_{1}, ~p_{2}$ are the probabilities per pump pulse to generate one and two photon pairs, respectively. We assume that the probability for generation of more than two photon pairs is negligible. As detailed in the previous sections, these photons are passed through a HWP and a PBS, assigned time-bin $\te$ and $\tl$ corresponding to $H$ and $V$ polarizations, respectively, by the delay line and finally recombined using a fiber beamsplitter. For simplicity, we consider the HWP angle to be $0^{0}$ so that the ideal state would be an antibunched state.  If the fiber coupling efficiency is $\eta$, the multi-photon state in the fiber is
\begin{eqnarray}
\nonumber  \lk \Psi \rk \simeq &&  \sqrt{\eta^{2} p_{1}} \lk 1_{e}, 1_{l} \rk +   \sqrt{\eta^{2} \left(1-\eta\right)^{2} p_{2}} \left( \lk 2_{e}, 0_{l} \rk  + \lk 0_{e}, 2_{l} \rk \right) + \\
  && \sqrt{2 \eta^{3} \left(1-\eta\right) p_{2}} \left( \lk 2_{e}, 1_{l} \rk + \lk 1_{e}, 2_{l} \rk \right) + \sqrt{\eta^{4} p_{2}} \lk 2_{e}, 2_{l} \rk.
\end{eqnarray}
Here, the state $\lk 2_{e}, 1_{l} \rk$ represents the case when there are two photons in the {\it early} time-bin and one photon in the {\it late} time-bin, and so on. Also, we have retained only those terms which have at least two photons and therefore, can lead to coincidence counts at the two detectors. Using this relation, we see that the probability of detecting two photons in the {\it early} or {\it late} time bins is
\begin{eqnarray}\label{Prob_EE}
\nonumber && p\left(e,e\right) = p\left(l,l\right) \\
&&= 2 \left( \eta^{2} \left(1-\eta\right)^{2} p_{2} + 2 \eta^{3} \left(1-\eta\right) p_{2} + \eta^{4} p_{2}\right) = 2 \eta^{2} p_{2}
\end{eqnarray}
and that for detecting one photon each in {\it early} and {\it late} time bins is
\begin{equation}
p\left(e,l\right) \simeq \eta^{2} p_{1}.
\end{equation}
The extra factor of two in eq.(\ref{Prob_EE}) is because of the beamsplitter used for JTI measurements. Therefore, the relative probability of bunched to anti-bunched photons is
\begin{equation}
\frac{p\left(e,e\right)}{p\left(e,l\right)} = \frac{2 p_{2}}{p_{1}}.
\end{equation}
In our experiment, the SPDC was pumped with 300 mW of power with $p_{1} \approx 0.1$ and $p_{2} = \frac{g_{2}\left(0\right)}{2} p^{2}_{1} \approx 0.009$, where $g_{2}\left(0\right) \approx 1.8$ is the second-order intensity correlation function at zero delay. Therefore, the probability of detecting bunched events to anti-bunched events, for an ideally anti-bunched two photon state, is $\approx 0.2$. This agrees well with the experimental observation in Fig.\ref{fig:3}(c) and (h). \\


\begin{acknowledgments}
This research was supported by AFOSR-MURI FA9550-14-1-0267, ONR, Sloan Foundation and the Physics Frontier Center at the Joint Quantum Institute. We thank T. Huber, J. Fan and G. Solomon for fruitful discussions, Q. Quraishi for kindly providing the nanowire detectors, A. Migdall and S. Polyakov for the HydraHarp time-tagging module and J. Bienfang for the high power amplifier.
\end{acknowledgments}


%

\end{document}